\documentclass[useAMS,usenatbib]{mn2e}
 \usepackage{txfonts}
 \usepackage{graphicx}
\usepackage[T1]{fontenc}

 \title[{\it X-Shooter} spectra and {\it RXTE} light curves of the UCXB candidate 4U~0614+091]
{Time-resolved {\it X-Shooter} spectra and {\it RXTE} light curves of the ultra-compact X-ray binary candidate 4U~0614+091\thanks{ Based on {\it X-Shooter} observations obtained under the GTO time proposal ESO 084.D-0814(B) }}

\author [Madej et al.]
{O.K. Madej$^{1,2}$\thanks{E-mail: O.Madej@sron.nl}, P.G. Jonker$^{2,1,3}$, P.J. Groot$^{1}$, L.M. van Haaften$^{1}$, G. Nelemans$^{1,4}$, \newauthor T.J. Maccarone$^{5}$\\
\\
\normalsize{$^{1}$Department of Astrophysics/IMAPP, Radboud University Nijmegen, P.O. Box 9010, 6500 GL Nijmegen, The Netherlands}\\
\normalsize{$^{2}$SRON Netherlands Institute for Space Research, Sorbonnelaan 2, 3584 CA Utrecht,The Netherlands}\\
\normalsize{$^{3}$Harvard-Smithsonian Center for Astrophysics, 60 Garden Street, Cambridge, MA 02138, USA}\\
\normalsize{$^{4}$Institute for Astronomy, KU Leuven, Celestijnenlaan 200D, 3001 Leuven, Belgium}\\
\normalsize{$^{5}$University of Southampton, School of Physics and Astronomy, Highfield Campus, Southampton, Hampshire, SO17 1BJ}}

\begin{document}

\date{}

\pagerange{\pageref{firstpage}--\pageref{lastpage}} \pubyear{-}

\maketitle
\label{firstpage}
\def\apjl{ApJ}
\def\aj{AJ}
\def\apj{ApJ}
\def\pasp{PASP}
\def\spie{SPIE}
\def\apjs{ApJS}
\def\araa{ARAA}
\def\aap{A\&A}
\def\nat{Nature}
\def\mnras{MNRAS}
\def\prd{Phys.Rev.D}

\def\lsim{\mathrel{\rlap{\lower4pt\hbox{\hskip1pt$\sim$}}
    \raise1pt\hbox{$<$}}}                
\def\gsim{\mathrel{\rlap{\lower4pt\hbox{\hskip1pt$\sim$}}
    \raise1pt\hbox{$>$}}}   

\begin{abstract}
In this paper we present {\it X-Shooter} time resolved spectroscopy and {\it RXTE PCA} light curves of the ultra-compact X-ray binary candidate 4U~0614+091. The {\it X-Shooter} data are compared to the {\it GMOS} data analyzed previously by \citet{Nelemans2004}. We confirm the presence of C III and O II emission features at $\approx4650$ \AA\ and $\approx$ 5000 \AA. The emission lines do not show evident Doppler shifts that could be attributed to the motion of the donor star/hot spot around the center of mass of the binary. We note a weak periodic signal in the red-wing/blue-wing flux ratio of the emission feature at $\approx 4650$ \AA. The signal occurs at $P=30.23\pm0.03$ min in the {\it X-Shooter} and at $P=30.468\pm0.006$ min in the {\it GMOS} spectra when the source was in the low/hard state. Due to aliasing effects the period in the {\it GMOS} and {\it X-Shooter} data could well be the same. We deem it likely that the orbital period is thus close to 30 min, however, as several photometric periods have been reported for this source in the literature already, further confirmation of the 30 min period is warranted. We compare the surface area of the donor star and the disc of 4U~0614+091 with the surface area of the donor star and the disc in typical hydrogen-rich low-mass X-ray binaries and the class of AM Canum Venaticorum stars and argue that the optical emission in 4U~0614+091 is likely dominated by the disc emission. Assuming that the accretion disc in 4U~0614+091 is optically thick and the opening angle of the disc is similar to the one derived for the hydrogen-rich low-mass X-ray binaries we conclude that the donor star may in fact stay in the disc shadow. Additionally, we search for periodic signals in all the publicly available {\it RXTE PCA} light curves of 4U~0614+091 which could be associated with the orbital period of this source. A modulation at the orbital period with an amplitude of $\approx 10$\% such as those that have been found in other ultra-compact X-ray binaries (4U~0513$-$40, 4U~1820$-$30) is not present in 4U~0614+091. We propose that the X-ray modulation found in 4U~0513$-$40 and 4U~1820$-$30 could be a signature of an outflow launched during the high/soft state near the stream impact region. 
\end{abstract}


\section{Introduction}
The source 4U~0614+091 is a persistent ultra-compact X-ray binary (UCXB) candidate that belongs to the class of low-mass X-ray binaries (LMXBs). UCXBs are characterized by very short orbital periods, typically less than 80 min. The consequence is a secondary star with such a high density that it excludes main-sequence type donor star \citep{Nelson1986,Savonije1986}. The donor star in 4U~0614+091 is most probably a CO white dwarf, based on the emission lines of oxygen and carbon found in the optical spectra and the lack of hydrogen and helium lines \citep{Nelemans2004,Werner2006}. An additional hint towards a CO white dwarf comes from the detection of the relativistically broadened O VIII Ly$\alpha$ emission line \citep{Madej2010, Schulz2010}. The accretor is a neutron star, since it is showing type I X-ray bursts \citep{Kuulkers2009}. The source shows no eclipses or dips in the light curves. However, due to a very low mass ratio (theoretical estimate based on the suggested orbital period of 50 min is $q\approx 0.01$) the threshold for the inclination below which the eclipses and dips are not present can even be higher than 80 deg \citep{Horne1985,Paczynski1974}. \\
\begin{figure*}  
\includegraphics[width=0.95\textwidth]{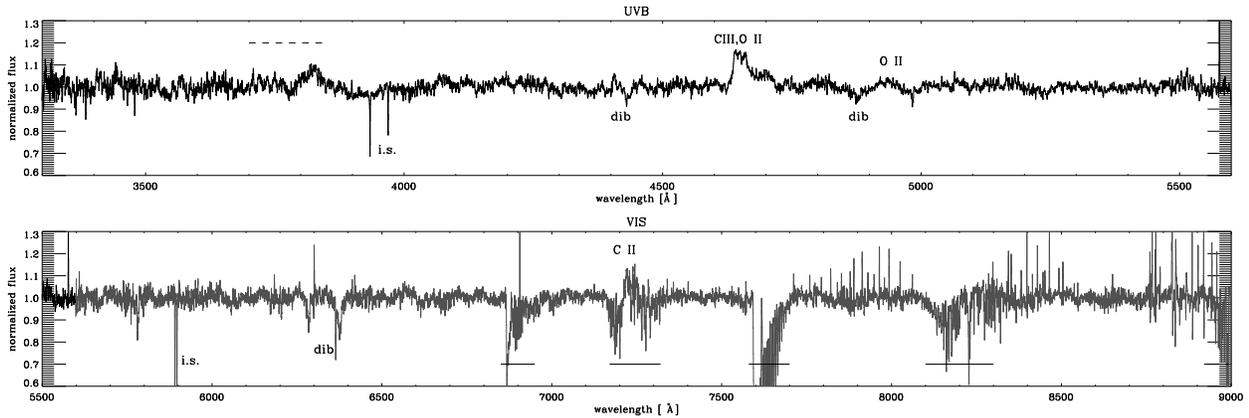}
 \caption{The average normalized {\it X-Shooter} spectrum of 4U~0614+091. Interstellar features, diffuse interstellar bands and telluric absorption lines are marked with 'i.s.', 'dib' and a horizontal line, respectively. {\it Top panel}: The average UVB arm spectrum. Note the emission feature at $\approx$ 4650 \AA\ which most probably is a blend of carbon and oxygen lines. The features around 3800 \AA, marked with a dashed line are artifacts and they occur due to the imperfect merging of the master flats. {\it Bottom panel}: The average VIS arm spectrum. The residuals from the sky subtraction become more prominent above 7000 \AA. } 
\end{figure*}
The orbital period measurement is important to classify the source as an UCXB and determine its evolutionary stage. In the case of 4U~0614+091 the first claim of the orbital period measurement of $\approx$ 50 min was by \citet{OBrien2005} in high-speed optical data taken with {\it ULTRACAM}. The modulation in the optical light curve can occur either due to heating of one side of the donor by X-ray irradiation, the superhump phenomenon caused by a tidal deformation of accretion disc, emission from the impact point of the stream with the disc or an asymmetry of the accretion disc. The orbital period of $\approx$ 50 min was also suggested by \citet{Shahbaz2008} in {\it IAC80} data and \citet{Zhang2012} in {\it Otto Struve Telescope} data. However, \citet{Hakala2011} analyzed a large set of fast optical photometry obtained with the {\it NOT} telescope and report no conclusive evidence for the orbital period of this system. In one out of 12 optical data sets the authors found a strong periodic signal at $\approx$ 50 min. \\
The modulations with the orbital periods in the UCXBs were reported not only in optical but also X-ray light curves \citep{Fiocchi2011, Stella1987}. As far as 4U~0614+091 is concerned a periodic modulation of $\approx$ 25 min was found in one {\it EXOSAT} light curve \citep{Hakala2011}, when the source was in a high flux state. On the other hand, \citet{Schulz2010} folded a {\it Chandra HETGS} light curve, when the source was in the low/hard state, with the periods of 41, 51, 62 and 120 minutes and found no evidence of a periodic modulation. \\
A powerful method of measuring the orbital parameters in persistent LMXBs or LMXBs in outburst is to measure the Doppler shifts of the high-excitation
emission lines of C III and N III forming the Bowen blend at $\approx4650$ \AA\ \citep{Steeghs2002}. These narrow lines are associated with the irradiated
companion star and have been detected in many LMXBs \citep{Casares2003,Casares2006,Cornelisse2007a,MunozDarias2009}. \\
Another method used to determine the mass ratio is to measure the rotational broadening of the absorption lines originating in the atmosphere of the donor star \citep{GIes1986}. \\
\citet{Nelemans2006} analyzed a set of time-resolved spectra of 4U~0614+091 taken with {\it Gemini Multi-Object Spectrograph} ({\it GMOS}) on the {\it Gemini-North} telescope and found no signatures of narrow emission lines. Although the source is showing a broad emission blend of O II and C III lines at $\approx$ 4650 \AA\ the resolution of $\approx$ 2 \AA\ might have been too low to detect narrow line components, if present. The authors found, however, weak evidence for a sinusoidal pattern with a period of 48.547 min in a weak absorption line around 4960 \AA\ (possibly caused by C I), which could result from the absorption in the atmosphere of the donor star. \\
A spectroscopic measurement of the orbital period has not been successful for any of the UCXBs except for the weak evidence noted by \citet{Nelemans2006}. It has been, however, successful for AM Canum Venaticorum (AM CVn) stars. This class of sources consists of a white dwarf accretor with a degenerate or semi-degenerate, hydrogen-poor donor star. They are similar to UCXBs in terms of the mass ratio and chemical composition of the donor star. The Doppler shifts of emission lines found in the optical spectra of AM CVns are caused by the motion of the impact point of the accretion stream onto the accretion disc \citep[e.g.][]{Nather1981,Roelofs2005} or, in the case of the very short period system HM Cnc, the irradiated side of the donor star \citep{Roelofs2010}. \\
In this paper we analyze time-resolved {\it X-Shooter} spectra of 4U~0614+091 searching for narrow emission/absorption lines showing periodic Doppler shifts associated with the donor star/hot spot motion around the center of mass. The {\it X-Shooter} resolution of $\approx$ 0.9 \AA\ in the blue part of the spectrum gives us a better opportunity to detect the narrow emission components than previous studies using the {\it GMOS} instrument \citep{Nelemans2006}. We compare the {\it X-Shooter} and the {\it GMOS} data. Additionally, we analyze all archival {\it RXTE PCA} light curves in search for an X-ray modulation at the orbital period. We separate high and low X-ray flux regimes of the source and look for periodic signals in both of them. 
\begin{table*}
\begin{center}
\caption{Log of the {\it X-Shooter} observations. The time column indicates the start time of the first and the last observation obtained during each night. The third column shows the number of exposures in each arm. The average and ranges of the seeing values as well as the airmass during the time when the observations were taken are indicated. The seeing was measured from the spectra in the wavelength range 4500-4520 \AA. The range of S/N in the single spectra is calculated at $\approx 4500$ \AA\ (UVB), $\approx 6500$ \AA\ (VIS). In the case of the NIR arm we provide the S/N of an average spectrum for each observing night, measured around $12500$ \AA.}
\begin{tabular}{lccccccc}
\hline
\hline
Date  &  UT Time & \# of exp.$^{*}$ & Seeing [$^{\prime\prime}$] aver., range & S/N UVB & S/N VIS & S/N (aver.) NIR & Airmass \\
\hline
16/02/2010 & 01:33:32-02:20:01 & 15 & 0.81, 0.73-0.96 & 3.8-4.3& 5.2-6.3& 1.2 & 1.21-1.26 \\ 
18/02/2010 & 00:38:18-02:30:10 & 30  & 0.92, 0.66-1.12 & 5.4-7.1& 6.9-9.8& 0.8 & 1.21-1.29 \\
20/02/2010 & 00:36:55-01:30:42 & 15  & 0.93, 0.79-1.04 & 4.0-7.2& 5.4-9.4& 1.3 & 1.21-1.21 \\
21/02/2010 & 00:29:50-01:23:37 & 15  & 1.03, 0.91-1.21 & 5.8-6.6& 7.4-8.7& 1.3 & 1.21-1.21 \\
22/02/2010 & 00:30:39-01:24:26 & 15  & 0.94, 0.85-1.05 & 6.0-6.9& 7.7-8.9& 1.5 & 1.20-1.22 \\
\hline

\end{tabular}\\
{\footnotesize $^{*}$ In the case of the NIR arm three exposures were taken and averaged during the time of one UVB and VIS exposure} 
\end{center}
\end{table*}
\section{Observations and data reduction}
\subsection{{\it X-Shooter} observations}
4U~0614+091 was observed with the {\it X-Shooter} instrument on 5 nights from 16-22 February 2010 (see Table.1). {\it X-Shooter} is an echelle spectrograph operating at the Cassegrain focus of {\it Unit Telescope 2} (Kueyen) on the {\it Very Large Telescope}. In a single exposure the instrument covers the spectral range from the UV to the near-IR K band in 3 arms \citep{Vernet2011}. The ultraviolet-blue (UVB) arm covers 3000-5595 \AA\ in 12 orders with a resolution of 5100 (for a 1$^{\prime\prime}$ slit width), the visual-red (VIS) arm covers the range 5595-10240 \AA\ in 14 orders with a resolution of 8800 (for a 0.9$^{\prime\prime}$ slit width) and the near-infrared arm covers 10240-24800 \AA\ in 16 orders with a resolution of 5300 (for a 0.9$^{\prime\prime}$ slit width). The slit widths described here were used during the observations. \\
The source was observed in the {\sc slit stare} mode (fixed position on the slit) in all three arms. The binning factor for the UVB and VIS arm spectra was 2 in the spatial direction and 2 in the spectral direction, the NIR arm spectra are by default not binned. The slow readout mode with high gain (100k/1pt/hg/2x2) was used for all the nights except the first night when the readout mode was fast with low gain (400k/1pt/lg/2x2). The exposure time  of every observation  was 180 sec in UVB and VIS arm.  Due to significant thermal background noise in the NIR arm instead of one 180 sec exposure, three exposures of 60 sec were taken and averaged in one FITS file. 
\begin{figure}  
\includegraphics[width=0.5\textwidth]{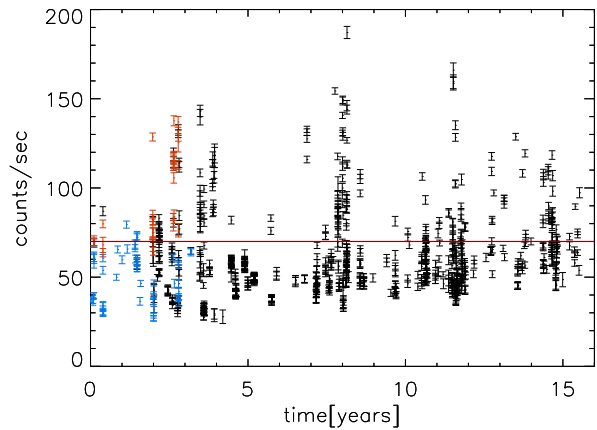}
\caption{Fluxes averaged over one observing ID of all {\it RXTE PCA, PCU 2} observations in the X-ray energy range 2-5 keV. The blue and orange measurements are taken from \citet{vanStraaten2000} and represent a sample of light curves in the island state up to lower banana state (blue) and lower banana up to higher banana state (orange). The red solid line at 70 cts/sec distinguishes roughly between the high/soft and low/hard state.}
\end{figure}
We reduce the data using the standard {\it X-Shooter} reduction pipeline version 1.5.0 \citep{Goldoni2006,Modigliani2010}. A physical model mode was used during the reduction process in which the solution is obtained by optimizing the instrument physical model parameters to the data. Since the data were obtained in February 2010 during the first, shared-risk semester of {\it X-Shooter} operations, no linearity frames are available for these observations, therefore we do not compute the detector gain and linearity ({\sc xsh\_lingain} recipe). The bad pixel map is included only during the NIR arm data reduction process, since it contains significantly more bad pixels than the UVB and VIS arm data. In the UVB arm we use an instrumental response updated by the {\it X-Shooter} calibration team which contains a more accurate than currently available in the pipeline flux calibration around $3600-3800$ \AA\ (Wolfram Freudling, private communication). In the VIS arm we use the instrumental response file provided with the pipeline. The calibration of the source spectrum is additionally corrected for the instrumental flexures during the observing nights using the {\sc xsh\_flexcomp} recipe. In order to estimate the sky background, a one-dimensional spectrum of the sky is built. The pixels outside of the source extraction region and far from the slit edges are taken. The interpolation of the resulting one-dimensional sky spectrum is done by applying the running median with the default window width of 14 pixels. The optimal extraction is used to obtain the one-dimensional spectrum of the source.   \\
The analysis of the spectra is performed using the {\sc molly} package developed by Tom Marsh. 
\subsubsection{Average spectrum}
We extract all UVB and VIS spectra from every night and calculate the weighted (based on S/N ratio) average UVB and VIS spectrum (see Fig. 1). The resolution in the output UVB and VIS spectra is $\approx$ 0.9 \AA. All the spectra are extracted with a wavelength sampling of 0.3 \AA\ and a median smoothing is used with a window width of 4 bins for UVB arm and 5 bins for VIS arm. The plate scale is $\approx 0.32$ arcsec/pix for UVB and VIS. We determine the seeing for every observation by measuring the FWHM of the source profile of each spectrum in the wavelength range 4500-4520 \AA\ (see Table 1; for the seeing measurement the spectra are unsmoothed and binned to the resolution of the instrument). The accuracy of the wavelength calibration in every spectrum is determined by measuring the position of the O I sky lines in all spectra. The sky line at 5577.338 \AA\ has an rms scatter of 0.05 \AA\ (UVB arm) and the sky line at 6300.304 \AA\ has an rms scatter of 0.07 \AA\ (VIS arm). \\
Since one of the goals of this paper is to look for periodic Doppler shifts in the emission lines that could correspond to the orbital period of this source, we examine the spectra `by eye' in search of prominent emission lines. We confirm the presence of an emission feature at $\approx$ 4650 \AA\ reported by \citet{Nelemans2004,Nelemans2006} and \citet{Werner2006} corresponding to the blend of C III and O II lines (see Fig. 1). We also find weak emission line signatures of O II at $\approx 4900$ \AA\ and C II at 7200 \AA\ \citep{Nelemans2006,Werner2006}. The combination of two different flat fields (D2 and QTH) causes a discontinuity in the response curve and results in the artifacts in the flux calibrated spectra around 3800 \AA\, hence this region is difficult to examine for possible emission features. Therefore, we check individual orders in search for emission lines of O III mostly \citep{Werner2006} around 3700 \AA. However, we find no significant emission features that could be attributed to the emission lines of oxygen or carbon in that region. The blue part of the UVB spectrum (3000-3600 \AA) shows small artifacts (`bumps' and `dips') visible also in the spectrum of the standard star possibly caused by imperfect merging of the spectral orders in the region where the S/N ratio of the data is much lower than in the middle and red part of the spectrum.\\
We extract the median NIR spectrum from all observations taken each night. The resolution of the NIR spectrum is $\approx$ 3 \AA\ and plate scale is $\approx 0.21$ arcsec/pix. We detect emission at wavelengths where the near-IR J, H and K bands occur. 
Most of the LMXBs are located towards the Galactic center. Therefore, their optical light is usually highly absorbed. A near-IR spectrum could prove an alternative way to study the chemical composition of the donor star of the UCXBs and distinguishing between the CO or He WD. For that reason we look for significant differences between the LTE model spectra for UCXBs with a carbon-oxygen and helium WD donor star \citep{Nelemans2006}. There is a strong emission line at $\approx$ 1.2$\ \micron$ (blend of C III transitions: 1199.13, 1198.8 and 1198.12 nm) present only in the CO model. Therefore, its presence could be used to distinguish an UCXB with a CO WD donor from He WD donor star. We examine the wavelength range where this emission line should occur. We find, however, no evidence for the presence of such a line. The upper limit on the flux of this line using the spectrum is also difficult to determine due to the systematic errors in the flux calibration. 
Due to a poor S/N ratio of the NIR spectra and many residuals from the sky lines subtraction we do not discuss this part of the spectrum further. 
\begin{table}
\begin{center}
\caption{The measured equivalent widths for the emission features: C III and O II transitions at 4624-4680 \AA\ and O II transitions at 4680-4720 \AA\ and 4900-4960 \AA.}
\begin{tabular}{lccc}
\hline
\hline
Date & EW (C III, O II)  & EW (O II)  \\
&(4624-4680 \AA) & (4680-4720 \AA,  \\
&& 4900-4960 \AA) \\
\hline
16/02/2010 & 6.7$\pm$0.1 & 4.0$\pm$0.2 \\
18/02/2010 & 3.25$\pm$0.06 & 1.83$\pm$0.09 \\
20/02/2010 & 4.67$\pm$0.09 & 2.4$\pm$0.1 \\
21/02/2010 & 5.59$\pm$0.1 & 2.8$\pm$0.1 \\
22/02/2010 & 4.6$\pm$0.1 & 2.5$\pm$0.1 \\
\hline
\end{tabular}

\end{center}
\end{table}
\subsection{{\it RXTE} observations}
We choose all of the Standard 2 mode (FS4a*.gz), PCU 2 data obtained during the {\it RXTE} mission ($\approx 16$ years) of 4U~0614+091. The observations where type I X-ray bursts were detected (2 light curves which contain tails of the X-ray bursts at MJD=51789 and 51944, \citealt{Kuulkers2009}) are excluded and the {\sc ftools} package is used to process the rest of the \textit{RXTE} data. The data are screened with standard criteria and the light curves are extracted using the {\sc saextrct} tool in the channel range  2-14 (roughly 2-5 keV \footnotemark) with time bins of 16 sec. In this way our analysis is more sensitive to mechanisms causing the modulation of the soft X-rays. The background light curve is created by the {\sc runpcabackest} tool using the background model for bright sources and subtracted from the source light curve using the {\sc lcmath} tool. We do not apply any correction for the deadtime. In total there are 488 {\it RXTE PCA} obsIDs and the length of the individual observations is less than 55 min. The majority of the observations have a length in the range 40-60 min which is similar to the orbital period proposed so far for 4U~0614+091. Hence if the orbital period of 4U~0614+091 is indeed $\approx50$ min we may expect prominent aliases in the periodogram which can make the detection of the true signal more difficult. \\
The light curves are analyzed using the {\sc Period} Time-Series Analysis Package available in {\sc starlink namaka} package.
\footnotetext[1]{The gain of the {\it PCA} instrument used to drift slowly with time, which slightly changed the energy to channel assignment. }
\begin{center}
\begin{figure}  
  
\includegraphics[width=0.5\textwidth]{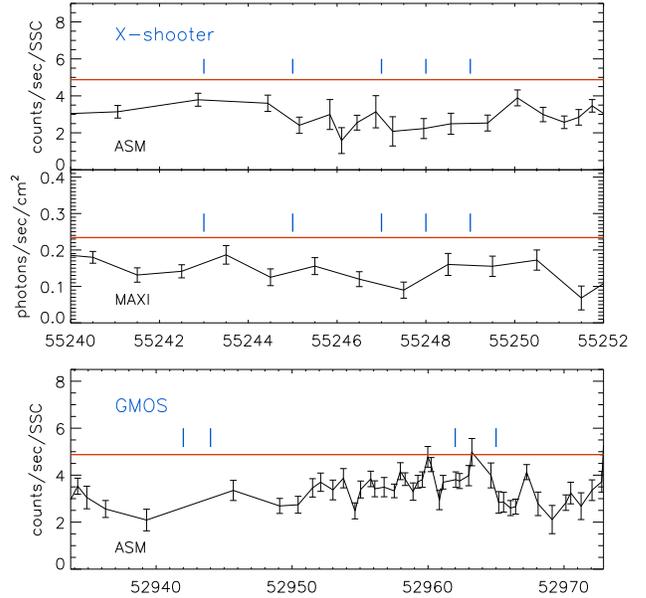}
\caption{{\it Top panel}: {\it RXTE ASM} and {\it MAXI} flux of 4U~0614+091 close in time to the {\it X-Shooter} observations. The red line shows the approximate threshold between the low/hard and high/soft state determined from the {\it PCA} fluxes (see Sec 3.4). The blue vertical lines indicate positions of the {\it X-Shooter} observations. A flux of 1 Crab=2.4$\times 10^{-8}$ erg/sec/cm$^{2}$ corresponds to $\approx 75$ cts/sec/SSC for the {\it ASM} and 3.6 ph/sec/cm$^{2}$ for the {\it MAXI} instrument. {\it Bottom panel:}. {RXTE ASM} flux of 4U~0614+091 close to the optical {\it GMOS} observations (blue vertical lines).}
  
\end{figure}
\end{center}
\begin{figure}
\includegraphics[width=0.5\textwidth]{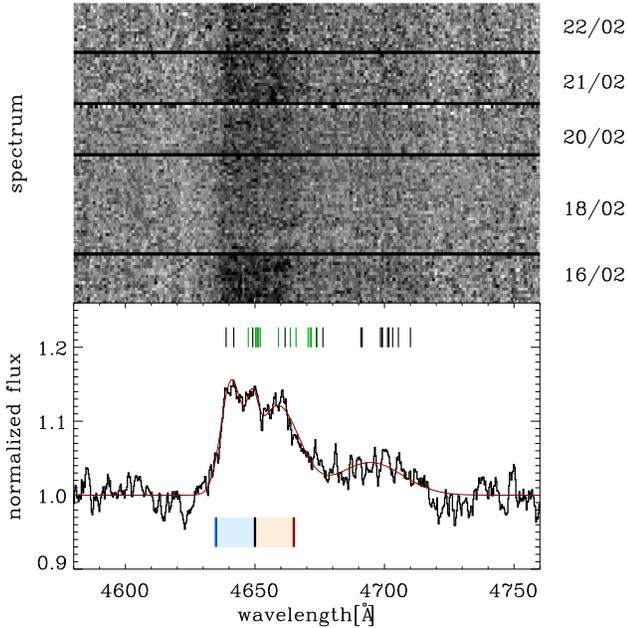}
\caption{{\it Upper panel}: Trailed spectrogram of all of the UVB observations showing the same blend of C III and O II as in the lower panel around 4650 \AA. Black stripes separate the data from each night. The width of the emission feature is $FWHM\approx$ 30 \AA\ $\approx$ 1910 km/sec and is consistent with being constant during the observations. {\it Lower panel}: The average UVB spectrum around $\approx$ 4650 \AA\ showing the emission feature which includes C III and O II transitions, most likely broadened by the motion in the accretion disc. The constant and four Gaussian lines fitted to the feature are overplotted (red line). The black and green vertical lines indicate O II and C III transitions, respectively, identified by \citet{Nelemans2004} in their best-fit LTE model. The light blue, light red areas below the spectrum indicate the flux range of the blue and red wing of the feature we use in the method to search for periodic signals, respectively. }
\end{figure}
\section{Analysis and results}
\subsection{Optical emission}
\subsubsection{The structure and variability of the emission features}
In order to investigate if there are any changes in the mean wavelength of the emission feature caused by e.g. changes in the ionization state between the observations, we cross-correlate the average spectrum from every observing night covering the emission feature at $\sim4650$ \AA\ with the average from all observing nights using {\sc xcor} in {\sc molly}. For the data analysis of the emission features at $\sim 4650$ \AA\ and $\sim 4690$ \AA\ we use the errors based on the variance of the data in the wavelength range 4500-4600 \AA. The cross-correlation reveals that the emission feature is stationary between the observing nights within the 1$\sigma$ errorbar calculated by the {\sc xcor} function. The width of the feature also does not change significantly between the observing nights.\\
We measure the equivalent widths of the emission features between $4624-4680$ \AA\ and $4680-4720$ \AA\ together with $4900-4960$ \AA. We note a significant decrease in the equivalent widths of the emission features between the first and second observing night which seems to correlate with the increase of the seeing value (see Table 1 and 2). Since the seeing conditions were variable over the observing nights and the signal to noise ratio of the spectra and emission features low it is difficult to say whether the variations in the equivalent widths of the emission features are caused by the intrinsic variations of the source emission. \\
In order to estimate the width and centroid of the emission features at $\approx 4650$ \AA\ and $\approx 4690$ \AA\ we first fit two Gaussian lines to the average spectrum from all observing nights which give a poor fit with reduced chi-squared value of $\chi^2/\nu=1476/360$. Additional visual inspection of the features lead us to fit four Gaussian lines (see Fig. 4, lower panel). The reduced chi-squared value of the fit using four Gaussian lines is $\chi^2/\nu=761/360$. The centroids and widths of the three distinct components forming the feature at $\approx 4650$ \AA\ are: $\lambda_{\rm cen1}=4640.7\pm0.1$ \AA,  $FWHM=9.3\pm0.3$ \AA; $\lambda_{\rm cen2}=4649\pm2.0$ \AA, $FWHM=5\pm4$ \AA; $\lambda_{\rm cen3}=4659\pm2$ \AA, $FWHM=20\pm4$ \AA. The centroid and width of the O II emission lines at 4680-4720 \AA\ are $\lambda_{\rm cen}=4694\pm2$ \AA\ and $FWHM=30\pm5$ \AA. \\

\subsubsection{X-ray flux during the optical observation}
We check the {\it RXTE ASM} and {\it MAXI} flux during the optical observation of the source (see Fig. 3). The flux is measured in the energy range 1.5-12 keV by {\it ASM} and we average 10 observations. In the case of the {\it MAXI} instrument we use the one day averaged flux in the energy range 2-20 keV. The red line indicates the approximate threshold between the low/hard and high/soft state based on the flux of the source (see Sec 3.4 for details) calculated for the energy range 2-20 keV (assuming power-law slope of $\Gamma=2$ and interstellar absorption of $N_{H}=3\times10^{21}$ cm$^{-2}$). The average X-ray flux level is around $10^{-9}$ erg/cm$^2$/sec (2-20 keV energy range) and indicates that 4U~0614+091 was in the low/hard state during the optical observations.
\subsection{Trailed spectra}
\citet{Nelemans2006} noticed possible variability in the weak absorption line around 4960 \AA\ (possibly C I at 4959 \AA) with a period of 48.547 min. Therefore we extract all UVB spectra separately and plot all the spectra in the form of a trailed spectrogram. We use the wavelength sampling of 0.3 \AA\ and median smoothing with a window of 4 bins. The normalization of the spectra is done by fitting a third order spline to the continuum of each spectrum with the most prominent carbon and oxygen emission lines masked. Each spectrum is then divided by the function resulting from the fit. \\ 
The expected radial velocity semi-amplitude of the primary is $K_1=5\times\sin i$ km/sec and that of the secondary is $K_2=727\times\sin i$ km/sec (assuming a 1.4 M$_{\odot}$ mass neutron star, 0.01 M$_{\odot}$ mass donor star and orbital period of 50 min). The maximum velocity broadening due to the integration time of a single exposure is $2\pi K_2\times T_{\rm exp}/P_{\rm orb}\approx$ $270$ km/sec. Here, $i$ is the inclination of the orbital plane with the plane of the sky.\\
We examine the region around 4959 \AA\ carefully and phase all of the spectra with periods in the range 48-49 min and step of 0.001 min. We find, however, no significant modulation such as that reported by \citet{Nelemans2006}.\\
\\
\begin{figure*}
\includegraphics[width=0.53\textwidth]{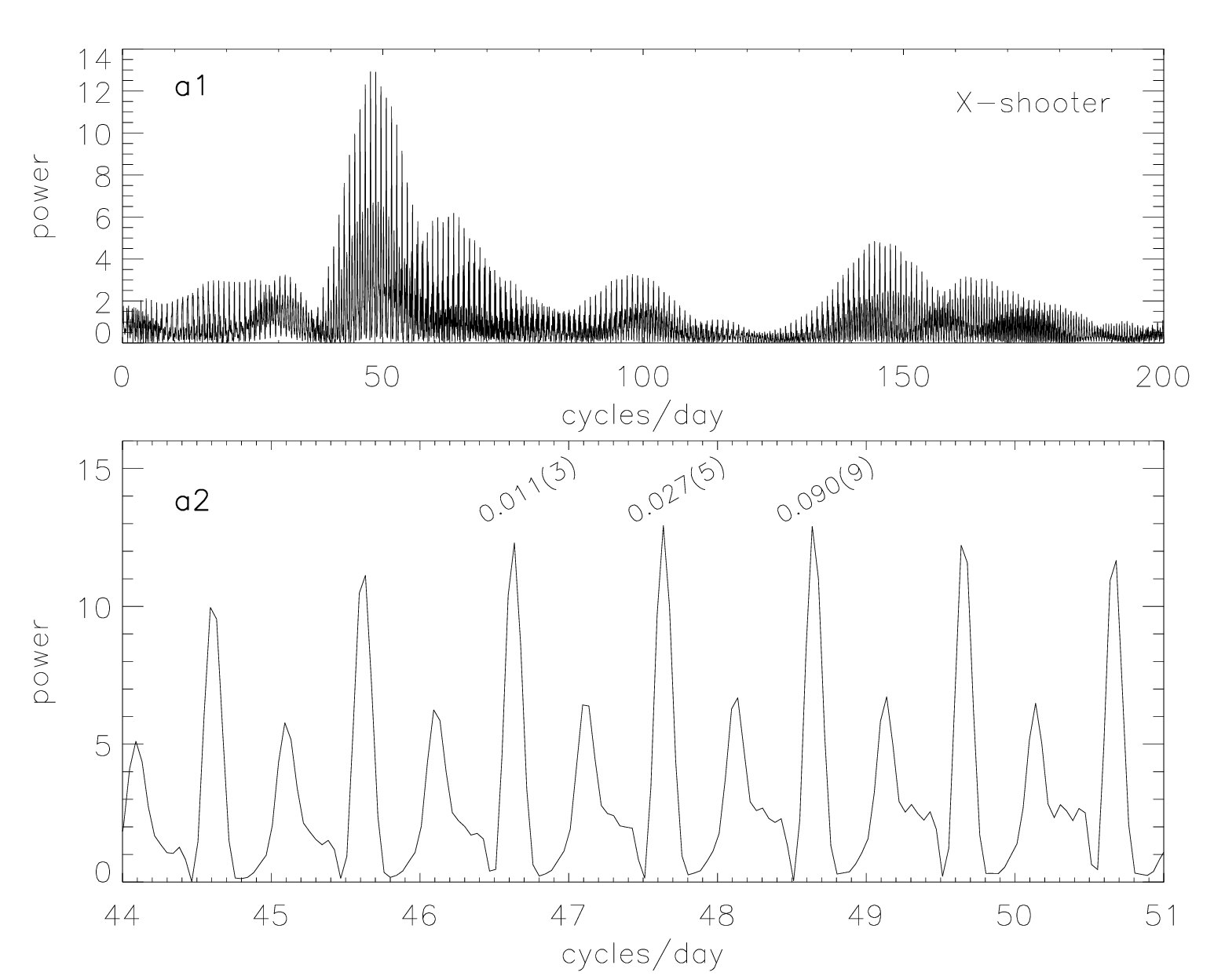}\includegraphics[width=0.52\textwidth]{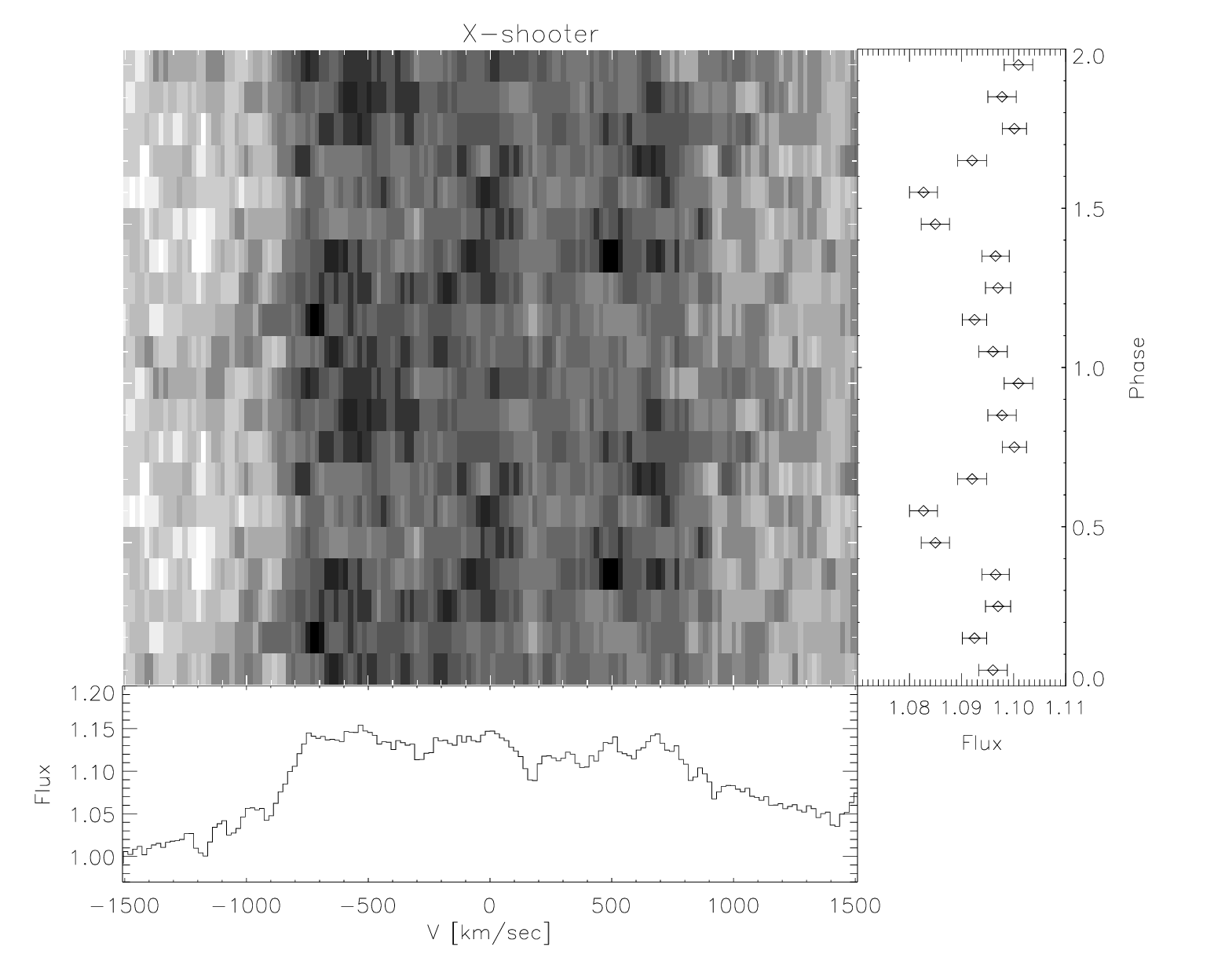}\\
\includegraphics[width=0.53\textwidth]{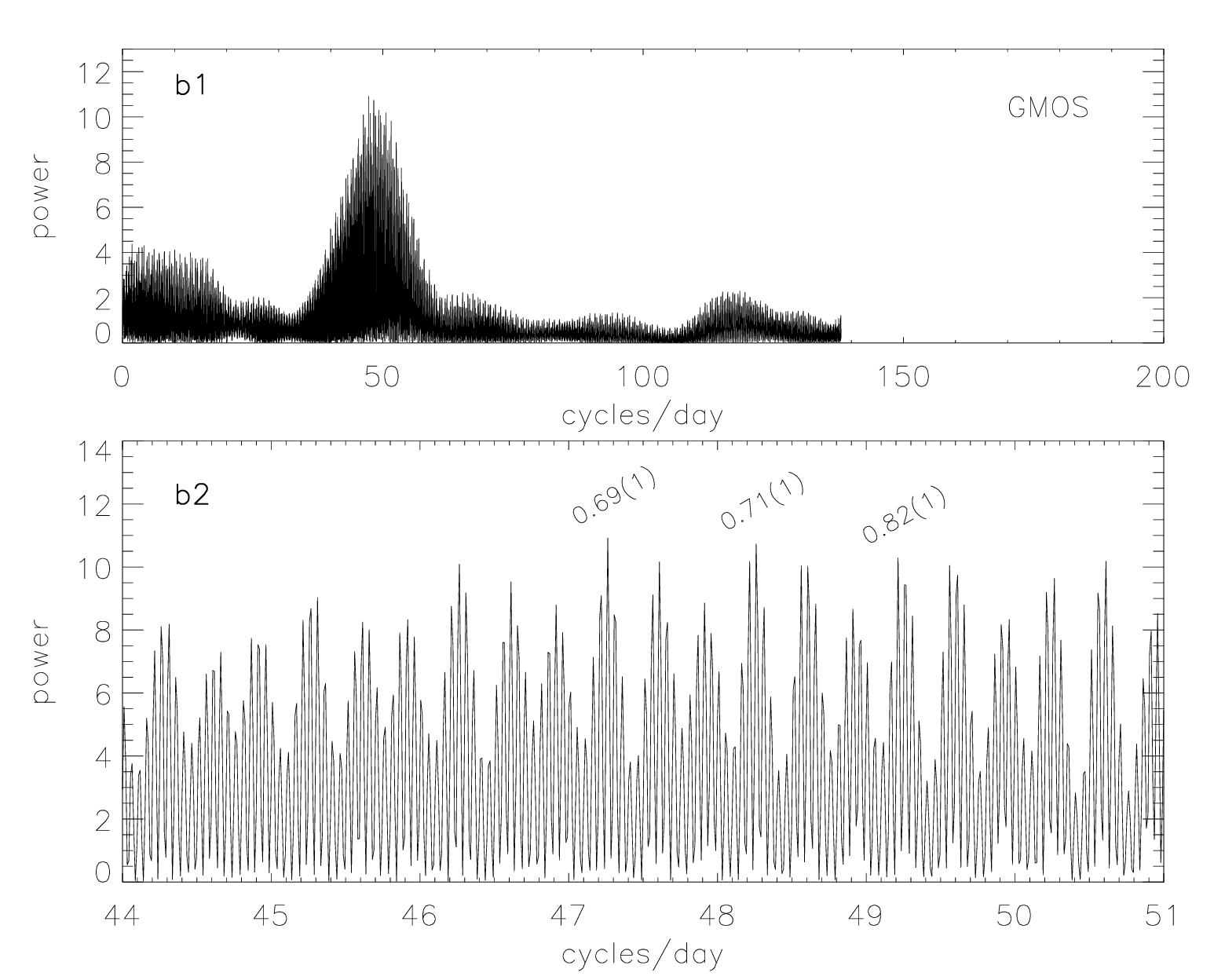}\includegraphics[width=0.52\textwidth]{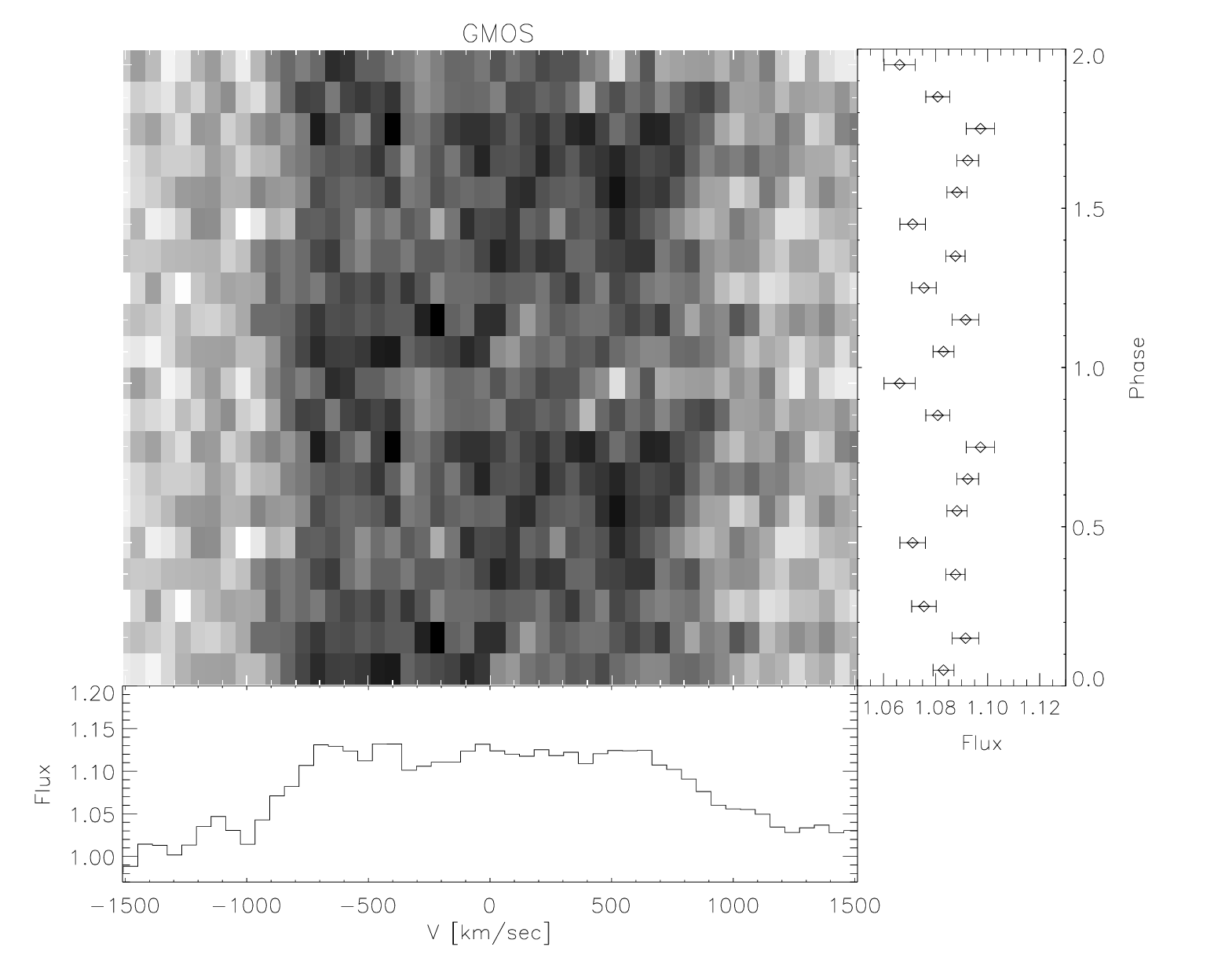}
\caption{{\it Left panel:} Lomb-Scargle periodograms using {\it X-Shooter} (a1 \& a2 panels) and {\it GMOS} (b1 \& b2 panels) data calculated from the flux ratio of the blue and red wing of the emission feature at $4650$ \AA\ measured in each individual spectrum. Note the peak at $\approx 50$ cycles/day $\approx$ 30 min in the {\it X-Shooter} (a1 \& a2 panels) as well as {\it GMOS} data (b1 \& b2 panels). The {\it X-shooter} periodogram is overplotted over the {\it GMOS} zoomed-in periodogram (orange curve, b2 panel). The $P2$ probabilities for the highest peaks in the periodograms and two aliases. {\it Right panel:} The trailed spectrogram constructed by folding the data on the period found in the Lomb-Scargle periodograms: $P=30.23\pm0.03$ min for {\it X-Shooter} data (upper panel),  $P=30.468\pm0.006$ min for {\it GMOS}. The spectrogram it plotted in velocity scale with one bin corresponding to 19.35 km/sec/px at 4650 \AA\ for {\it X-Shooter} and 60.43 km/sec/px at 4650 \AA\ for {\it GMOS}. Two periods are shown in the spectrograms. Note a weak periodic signal in the left, middle and right side of the feature. The average spectrum and average flux at $\approx 4650$ \AA\ is shown in the panel below and on the right side of the trailed spectrograms, respectively. }
\end{figure*}
A useful method to find the orbital period in time-resolved spectra is to look for the periodic changes in the flux ratio of the blue and red wing of the emission lines \citep{Nather1981}. It has been successfully applied to AM CVn sources \citep[e.g.][]{Roelofs2005}. In our case most emission features are not single transitions of e.g. helium as is the case in AM CVns but blends of oxygen and carbon transitions (see Fig. 4, lower panel). Therefore the periodic signals detected using this method could be caused by the Doppler shifts of the emission lines as well as the variable flux ratios between the carbon and oxygen emission lines forming the feature. \\
We apply this method to the most prominent feature at $\approx 4650$ \AA. 
We measure the ratio $R$ between the blue and red wing of the feature $R={\rm Flux}(4635-4650 \AA)/{\rm Flux}(4650-4665 \AA)$ for every observation and calculate the Lomb-Scargle periodogram. The periodogram shows a maximum power at $P=30.23\pm0.03$ min (see Fig 5., left panel). We note that choosing a different binning of the data (e.g. 0.4 \AA) and a different window width for the median smoothing (e.g. 3 bins) causes the highest peak in the periodogram to shift to the position of the closest alias (e.g. $P=29.61\pm0.03$ min). Hence it is likely that the true error on the detected period is larger than the quoted value.\\
The significance of the period is calculated using a Monte Carlo method with the number of permutations set to 1000. A Fisher randomization \citep{Nemec1985} is used to determine the significance of the peak and two significance estimates are given in the output. The first is {\it P1} and represents the probability that, given the frequency search parameters, no periodic component is present in the data with this period. The second significance {\it P2} represents the probability that the period is not equal to the quoted value but is equal to some other value. In our calculation the $P1\approx0.0$ ($P1$ lies between 0.00 and 0.01 with 95\% confidence level) and $P2=0.027\pm0.005$, which indicates that this period may well not be the true period.\\
Next, we create a  trailed spectrogram with spectra folded on an orbital period of 30.23 min. For this calculation the spectral wavelength scale is converted into a velocity scale with $\Delta\upsilon=19.35$ km/sec/px at 4650 \AA\ using the {\sc vbin} function in {\sc molly}. The heliocentric correction is additionally applied to all of the analyzed spectra. The trailed spectrogram reveals a very weak periodic pattern (see Fig. 5, right panel) on the left, middle and right side of the emission feature. The pattern on the blue side seems to be in antiphase with respect to the pattern in the middle and on the red side. Looking at the LTE model fitted to the emission feature \citep[see Fig. 4, ][]{Nelemans2004} the structure on the left side consists of the O II lines whereas the structure in the middle and right side is dominated by the C III lines. \\
Additionally, we create trailed spectrograms between 28-32 min with the step of 0.001 min. Since there are 10.000 frames to examine we create a movie ordering the frames with increasing orbital period. We notice a repeatable pattern separated by 1 cycle/day which correspond to the aliases seen in the Lomb-Scargle periodogram. Owing to the weakness of the signal it is difficult to attribute the periodic signal found to the orbital period of the source. \\
We investigate also the range of periods between 45-55 min by creating a number of trailed spectrograms with the step of 0.001 min. None of the frames, however, show evidence for a periodic signal that could correspond to the orbital periods suggested so far for 4U~0614+091 \citep[e.g.][]{Shahbaz2008}.
\subsection{Comparison between {\it X-Shooter} and {\it GMOS} data of 4U~0614+091} 
We reanalyze time-resolved spectra of 4U~0614+091 obtained by the {\it GMOS-North} \citep[see ][]{Nelemans2006}. In total there are 52 spectra taken between 30/10/2003 and 22/11/2003. The spectra are normalized and the wavelength scale is converted into velocity scale with  $\Delta\upsilon=60.43$ km/sec/px at 4650 \AA. We calculate the Lomb-Scargle periodogram from the flux ratio of the blue and red wing of the emission feature at $\approx4650$ \AA\ in the same way as for the {\it X-Shooter} data. The periodogram shows maximum power at $P=30.468\pm0.006$ min (see Fig 5., left panel) with probabilities $P1\approx0$ and $P2=0.69\pm0.01$. The detected period is close to the one detected in {\it X-Shooter} data but not the same. We calculate the $P2$ probabilities for the three highest peaks in the {\it X-shooter} and {\it GMOS} periodograms (see Fig 5., left panel). The $P2$ probability has a lower value for the alias on the left hand side of the highest peak in the {\it X-shooter} periodogram which indicates that the value of the period is more uncertain than indicated by the error on the highest peak. A high values of the $P2$ probability for all three peaks in the {\it GMOS} periodogram indicates that the detected period is uncertain as well. \\
Additionally, we create a trailed spectrogram by folding {\it GMOS} spectra with the detected period (see Fig 5., left panel). The pattern in the trailed spectrogram seems to be similar to the one found in {\it X-Shooter} data. 
\subsection{{\it RXTE} light curves: modulation of the X-ray emission} 
First, we consider all available {\it RXTE} data. In order to normalize the light curves we divide each by its average flux. Since the light curves are not evenly spaced in time, we use the Lomb-Scargle method to search for periodic signals. We calculate a single Lomb-Scargle periodogram taking the 16 year long dataset. The periodogram is calculated in the frequency range $10^{-4}-3\times 10^{-3}$ Hz (6-167 min). We choose a frequency resolution of $\Delta f=10^{-9}$ Hz based on the overall (16 years) time interval of the data. \\
The Lomb-Scargle periodogram reveals no clear periodic signal in the considered frequency range. The highest peak occurs at $\approx$ 84.9 min, which does not correspond to any orbital period proposed so far. Folding the data with this period gives a periodic signal with an amplitude of around 0.5\%. We note that this periodic signal is close to the orbital period of the {\it RXTE} satellite \citep{Wen2006}. 
Since the X-ray modulation found in the UCXB 4U~0513$-$40 was detected in the high/soft state of the source, we divide the light curves into a high flux group with {\it PCA} 2-5 keV count rate >  70 cts/sec which indicates a high/soft state and low flux group with {\it PCA} count rate < 70 cts/sec which indicates a low/hard state. This rough division between the island (low/hard state) and banana state (high/soft state) is done based on the results obtained by \citet{vanStraaten2000} concerning the correlation between the spectral shape and the timing features (indicating the state of the source) and source flux. The dataset analyzed by \citet{vanStraaten2000} covers observations performed during the years 1996-1998. In Fig. 2 we mark in blue those observations that are in the island state based on the spectral behavior and in orange those that are in the banana state. We calculate the Lomb-Scargle periodogram for the low and high flux data sets. There is again no clear evidence of a periodic signal in the light curves covering separately high/soft and low/hard state neither around $\approx 50$ min nor around $20-30$ min. The highest peak in the periodogram occurs at $\approx$ 81.5 min in the low/hard state and $\approx$ 78 min in high/soft state. Folding the data with this period gives a periodic signal with an amplitude of around 0.5\% and 1.3\% for the low/hard and high/soft state, respectively.

\section{Discussion}
We have analyzed {\it X-Shooter} data and reanalyzed {\it GMOS} data of 4U~0614+091 and find no clear signature of the orbital period around 50 min suggested by \citet{Shahbaz2008} and \citet{Nelemans2006}. We notice a weak periodic signature in the {\it X-Shooter} and {\it GMOS} data around 30 min, which could well be due to the orbital period. On the basis of the strong aliases present around the most probable peaks in the {\it X-shooter} and {\it GMOS} periodograms we cannot exclude that the two periods are consistent with being the same. It is possible that the modulation is caused by the Doppler shifts of the C III and O II lines as well as variable flux ratio of the C III with respect to the O II lines forming the feature. Perhaps we are observing two regions characterized by different ionization states (e.g. disc and a stream impact region). In order to confirm this interpretation, however, spectroscopic data with a resolution higher than {\it X-shooter} in combination with higher effective area would be necessary in order to resolve the narrow components and obtain sufficient signal to noise ratio.\\
Although there have been claims of the orbital period $\approx$ 50 min using optical data \citep{Shahbaz2008,Zhang2012} the authors found also indication for other periodic signals. \citet{Shahbaz2008} reports two periodicities at 64.1 min and 42 min in the {\it NOT} and {\it SPM} light curves, respectively. \citet{Zhang2012} finds a possibly quasi-periodic signal also at 16.2$\pm$0.1 min using data obtained by the {\it Otto Struve Telescope}. Therefore further confirmation of the signal at $\approx$ 30 min is warranted given the large number of reported periods and quasi-periodicities for this source.  \\

The lack of helium lines in optical spectra suggests that the donor star in the source is either a CO white dwarf or He star which is very late in its evolution, on the (semi-)degenerate branch of the evolutionary track \citep{Yungelson2008}. However, considering the bolometric flux of 4U~0614+091 estimated from the {\it RXTE ASM} data the source appears too bright to have a degenerate donor \citep{vanHaaften2012b}. The flux would match the theoretical predictions only if the orbital period of this source was around 25 min \citep{vanHaaften2012b}, which is more in line with the value detected in the {\it X-shooter} and {\it GMOS} data than the claimed 50 min period. \\
\begin{figure}  
\includegraphics[width=0.5\textwidth]{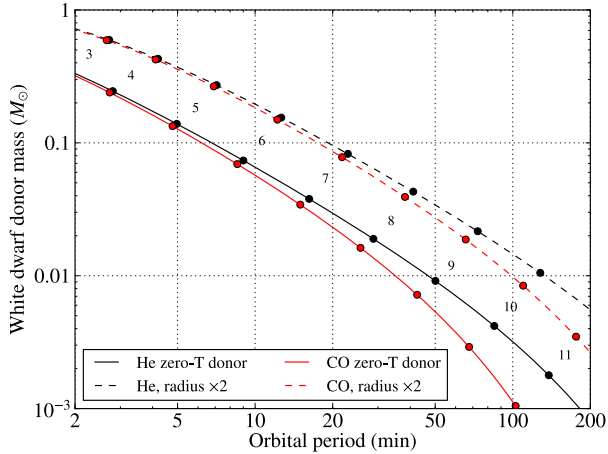}
\caption{The theoretical donor star mass as a function of the orbital period for He (black) and CO WD (red). The solid curve represents a zero-temperature track which neglects the thermal pressure in the WD, whereas the dashed curve illustrates an upper limit on the donor mass in the case a very high thermal pressure is introduced. The dots indicate the logarithm of the age calculated since the onset of mass transfer to the neutron star.} 
\end{figure} 
\subsection{X-ray reprocessing: 4U~0614+091 compared to LMXBs and AM CVns}
Since we find no clear evidence for Doppler shifts in the velocities of emission lines produced on the X-ray heated hemisphere of the donor star we consider the possibility that the optical emission, including the emission lines, is dominated by reprocessing of X-rays and the contribution from the donor star to the total as well as the reprocessed emission is negligible. \\
At first we estimate the visible area of the donor star and the disc as a function of the inclination. In this simple approach we assume that the ionization state and vertical structure of the accretion disc in hydrogen-rich LMXB and UCXB is similar. The shadowing of the donor by the disc is not included in this calculation, however, we discuss this problem later in this section. We use the formula describing the ratio between the donor and the accretor Roche lobe radius derived by \citet{Eggleton1983}. We estimate the projected area as a function of the inclination of the system assuming that the disc fills the entire Roche lobe of the primary 
\begin{equation}
S_{\rm frac}=\frac{S_{\rm WD}}{S_{\rm disc}+S_{\rm WD}}=\frac{0.5\pi R_{WD}^2 (1+\sin i)}{\pi R_{\rm disc}^2\cos i+0.5\pi R_{\rm WD}^2(1+\sin i)}
\end{equation}
$S_{\rm frac}$ is the ratio between the projected area of the WD ($S_{\rm WD}$) and the sum of $S_{\rm WD}$ and the projected area of the disc ($S_{\rm disc}$). We take into account in this calculation only the side of the donor star which is illuminated by the X-rays (e.g. for the inclination of 90 deg half of the donor star total surface is illuminated and visible for the observer). The surface area of the donor star and the disc to first approximation depends
on the mass ratio between the donor and accretor. 
In the case of the neutron star we assume the canonical mass of $M=1.4M_{\odot}$. We calculate the mass of CO and He WD donor stars as a function of the orbital period \citep{vanHaaften2012a,vanHaaften2012b} assuming that the thermal pressure in the WD is zero and the only forces present are the degeneracy pressure and the Coulomb interaction (Fig. 6, zero-temperature donor track). The Coulomb interaction becomes important in very low-mass WDs (< $10^{-3}$ $M_{\odot}$). In order to give an idea what would be the donor star mass in the case where the thermal pressure is non-negligible we calculate the donor star mass with a radius larger by a factor of 2. \citet{deloye2003} estimated that the influence of the thermal pressure on the radius of the white dwarf is probably around few tens of a percent. Hence, we should expect the donor star mass to be in between the solid and dashed curve (Fig. 6) and likely closer to the solid curve. \\
Fig. 7 shows the resulting surface area of the white dwarf secondary divided by the total surface area (donor star and the disc) as a function of the inclination (the mass ratio in 4U~0614+091 is assumed to be $q=0.01$). If the inclination of 4U~0614+091 is less than around 60 deg then the surface area of the donor with respect to the total surface area is significantly smaller for this source than for other, hydrogen-rich, LMXBs which show emission lines formed in the irradiated part of the donor star (see Table 3). Hence, the visible area of the donor and the disc in an UCXB with a low mass donor star and low inclination (less than around 60 deg) argues in favor of disc-dominated optical emission. \\  
The majority of the UCXBs found so far have orbital periods around 40-60 min \citep{intZand2007}, similar to that suggested for 4U~0614+091. The inclination of many is not well constrained. For most of them an upper limit is only given based on the lack of dips or eclipses in the light curve. However, since UCXBs likely have a lower mass ratio than hydrogen-rich LMXBs the inclination can reach values even higher than 80 deg without the occurrence of an eclipse \citep{Paczynski1974,Horne1985}. On the other hand the probability of having systems with a very high inclination is quite low (see Fig. 7, top axis). \\
Therefore, considering the difference between the visible fraction of the donor star and the disc surface in these UCXBs with respect to the LMXBs we may suspect the optical emission of UCXBs similar to 4U~0614+091 to be dominated by the disc emission. \\
\begin{figure*}  
\includegraphics[width=0.7\textwidth]{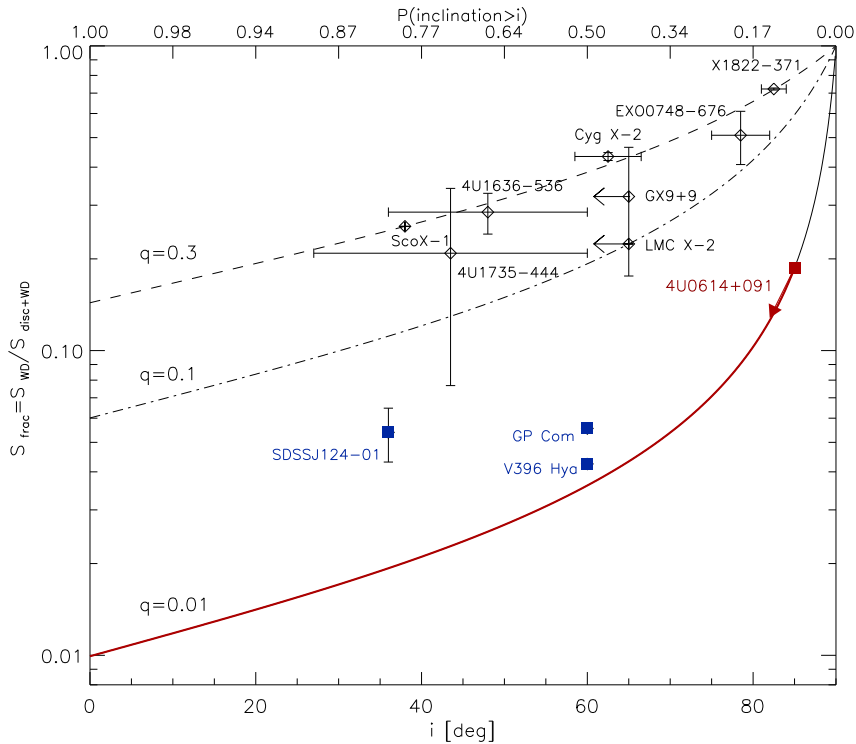}
\caption{The surface area of the donor star divided by the total surface area (donor star+accretion disc) as a function of binary inclination (see Eq. 1), overplotted are the best-estimate values for a number of LMXBs. Note that 4U~0614+091 could have a significantly lower donor star surface area compared to the total surface area than hydrogen-rich LMXBs for which the reprocessing of X-ray light in the donor star has been observed (red line). We plot (in blue) the positions of those AM CVns that have a mass ratio close to  4U~0614+091 and spectroscopic measurement of the orbital period. In the case we do not have the measurement of the inclination the source is plotted at 60 deg. The top axis represents the probability that the inclination of the system is higher than given for a homogeneous distribution in $\cos i$. } 
\end{figure*} 
\begin{table}
\begin{center}
\caption{A list of LMXB and AM CVn sources for which the mass ratio was measured from spectroscopy. The inclination (if constrained) and the reference for each source is also provided. In the case of AM CVn we list only the sources with mass ratio close to that of 4U~0614+091. }
\begin{tabular}{lccc}
\hline
\hline
Name  &  mass ratio & inc. [deg] & Reference \\
\hline
\multicolumn{4}{|c|}{{\bf LMXB}} \\
\hline
Sco X-1 & 0.3 & 38$$ & \citet{Steeghs2002}\\
X 1822-371&0.296-0.315&$82.5\pm1.5$& \citet{Casares2003}\\
4U 1636-536 & 0.21-0.34&36-60& \citet{Casares2006}\\
4U 1735-444 & 0.05-0.41 & 27-60& \citet{Casares2006}\\
EXO 0748-676 & 0.11-0.28 & 75-82& \citet{MunozDarias2009}\\
Cyg X-2 & 0.34$\pm0.02$ & 62.5$\pm4.0$ & \citet{Casares2010}\\
LMC X-2 & <0.1 & <65 & \citet{Cornelisse2007a}  \\
GX 9+9 & 0.07-0.35 & <65 & \citet{Cornelisse2007b}  \\ 
\hline
\multicolumn{4}{|c|}{{\bf AM CVn}} \\
\hline
GP Com &0.018&-& \citet{Marsh1999}\\
SDSS J1240$-$01 &0.039&36,53&\citet{Roelofs2005}\\
V396 Hya & 0.0125 & - & \citet{Ruiz2001}\\
\hline
\end{tabular}
\end{center}
\end{table}
As noticed earlier, the above calculation does not take into account the possible `shadowing effect' of the disc on the donor star. Hence, we estimate the height of the outer part of the disc and compare it with the theoretical estimates of the donor star radius. For a typical LMXB a half-opening angle of the disc is $\Theta\approx12$ deg \citep{deJong1996}. The carbon-oxygen disc could, however, be thinner by a factor of $\approx 2$ than a disc with solar composition in a typical LMXB \citep[whose prediction is based on the $\alpha$-disc theory]{Dunkel2006}. The height of the outer part of the disc for a disc radius of 0.45 $R_{\odot}$ is therefore in the range 0.045-0.09 $R_{\odot}$. The radius of the donor star assuming the zero-temperature track is $\approx 0.04\ R_{\odot}$ which is still smaller than the lower limit on the height of the outer part of the disc. In the case in which the radius is twice the radius expected for the zero-temperature track (very high thermal pressure) it is possible that the donor is being illuminated. Such a case is, however, unlikely given the typical influence of the thermal pressure on the radius of the white dwarf \citep{deloye2003}. \\
Apart from the geometry of the accretion disc we consider also its optical thickness. The donor star could still be illuminated by the X-rays that are passing through the accretion disc. \citet{Dunkel2006} have shown, however, that in the case of accretion disc dominated by ions of heavier elements such as helium, carbon or oxygen the electron and ion density as well as optical thickness of the disc increases (by a factor of few) with respect to the optical thickness of the disc in a typical hydrogen-rich LMXB. \\
Based on the theoretical predictions presented above it seems likely that the donor star is shielded from the radiation of the central source. If that is the case then any modulation of the optical light is unlikely to be caused by the reprocessing of the X-ray light in the donor star as is the case for typical hydrogen-rich LMXBs. \\
Fig. 7 indicates as well the position of a few AM CVn source for which the orbital periods were measured and the mass ratios are close to the predicted value in 4U~0614+091. Unlike in AM CVns, the accretion discs in LMXBs are strongly X-ray irradiated and the amount of reprocessing usually determines the temperature and elements visible in the optical spectra. Although the surface area of the donor star comparing to the accretion disc is similar in AM CVn source and UCXBs we may expect that the conditions in which the lines are formed in 4U~0614+091 are substantially different from those seen in AM CVn sources.\\
\subsection{X-ray modulation}
The origin of the X-ray modulation in the high/soft state that was observed for instance in the UCXB 4U~0513$-$40 \citep{Fiocchi2011} or UCXB 4U~1820$-$30 \citep{Stella1987} has not yet been established. In the case of the 4U~0513$-$40 a periodic modulation is visible when the source is in the high/soft state \citep{Fiocchi2011}. In 4U~1820$-$30 a periodic modulation with an amplitude of $\approx 10$\% was found when the source was in the high/soft state. The modulation was also observed when the source was in the low/hard state, however, the significance of the detection is reduced due to the factor of 3 lower count rate and factor of 2 lower modulation amplitude \citep{Stella1987}. According to the model of reprocessing presented by \citet{Arons1993} which describes the observations of UCXB 4U~1820$-$30 the X-ray light is reprocessed into the optical and ultra-violet light mostly in the outer regions of the optically thick, geometrically thin accretion disc. Additional 5$-$10\% of the light comes from reprocessing by the degenerate donor and is causing a prominent modulation with orbital period of the optical and ultra-violet light in 4U~1820-30. However, the X-ray light reflected off the surface of the donor star cause a modulation with an amplitude of only a few tenths of a percent of the total X-ray flux, hence it is insufficient to explain the large observed amplitude of the X-ray modulation in 4U~1820$-$30. \\
We propose that the modulation found in 4U~1820$-$30 and 4U~0513$-$40 may indicate the presence of an outflow launched during the high/soft state. The winds in several LMXBs are azimuthally symmetric and have an equatorial geometry \citep[see][]{Ponti2012}. However, in the case of an UCXB the outer radius to launch a thermal wind is comparable to the outer disk radius ($\approx$ $10^9$ cm, \citealt{Miller2006}). Therefore, when the point of impact of the gas stream from the companion occurs in the region where the wind is generated, the azimuthal symmetry is violated and more wind comes off this impact region rather than off other regions in the disk. Hence the foreground absorption will depend on the orbital phase. Additionally, since the wind in an UCXB is hydrogen deficient and helium or carbon and oxygen rich, the amount of absorption may be higher due to the strong dependence of the photoelectric absorption cross-section upon the atomic number. \\
A possible outflow present in 4U~1820$-$30 in the high/soft state has been reported by \citet{Costantini2012}. The authors found highly blue-shifted absorption lines of mildly ionized oxygen. The observed lines could originate in the stream impact region which could be less ionized than the accretion disc \citep[e.g.][]{vanPeet}. \\
In order to study a dependence of the neutral and ionized absorption on the orbital period using current grating instruments a long exposure time ($\approx$days) is required. Taking into account the fact that the observed variability of the UCXBs occurs on a time scale of days this study will be more feasible using future higher-effective area instruments like micro-calorimeter on {\it Astro-H} satellite. \\
Lack of a convincing periodic signal in the X-ray light curve with amplitude of few percent in the high/soft state of 4U~0614+091 indicates that the mechanism causing the modulation of X-rays in 4U~0513$-$40 is not present in 4U~0614+091. On the other hand the {\it RXTE/PCA} energy range, which does not cover the soft part of the spectrum  ($< 2$ keV), could be the reason why the modulation related to photoelectric absorption (most prominent in the soft part of the spectrum) is not detected. Alternatively a lower inclination of 4U~0614+091 with respect to the UCXBs showing the X-ray modulation could also explain the lack of the periodic signal with such a high amplitude. \citet{Ponti2012} demonstrate that the winds observed in a sample of black hole X-ray binaries in high/soft state have opening angles of few tens of degrees since they are only observed in dipping sources in which the disc is inclined at a large angle to the line of sight ($i \gtrapprox 60$ deg). A high-resolution $LETGS$ spectrum of 4U~0614+091 taken during the high/soft state does not show signatures (narrow absorption/emission lines) of an outflow \citep{Paerels2001}. Therefore, assuming that the UCXBs share the outflow properties with the sample of black hole X-ray binaries analyzed by \citet{Ponti2012} the inclination of 4U~0614+091 could in fact be lower than the threshold determined by the lack of eclipses in the X-ray light curve (see Sec. 4.1).\\
The limit on the amplitude of any X-ray modulation in the {\it RXTE} data of 4U~0614+091 is such that reflection of X-ray light off the surface of the donor star could still be below the detection threshold. If we assume that the star atmosphere reflects a fraction $\approx 0.5$ of the incident X-rays \citep{Anderson1981,London1981} then the amplitude of the X-ray modulation is expected to be lower than 1\% \citep{Arons1993} for the case of an UCXB. However, the theoretical prediction seems to indicate that the disc is likely blocking the X-rays from illuminating the donor star. 
\section{Conclusions}
We have analyzed time-resolved {\it X-shooter} and {\it GMOS} spectra of 4U~0614+091. We find a weak periodic signal at $\approx 30$ min in the red-wing/blue-wing flux ratio of the most prominent emission feature at $\approx 4650$ \AA. The modulation could be caused by the Doppler shifts of the C III and O II lines as well as variable flux ratio of the C III with respect to the O II lines forming the feature. Comparing the surface area of the donor and the disc of 4U~0614+091 with the surface are of the donor star and the disc in typical hydrogen-rich LMXBs and AM CVn sources, we conclude that the emission likely originates in the accretion disc. It is possible that this periodic signal represents the orbital period of the source, however, due to large number of periods reported for this source a further confirmation is needed.\\ 
Additionally we find no evident periodic signal which could be attributed to the orbital period of 4U~0614+091 in the {\it RXTE PCA} light curves covering the energy range 2-5 keV. 
\section{Acknowledgments}
We thank the referee for the useful comments. OKM \& PGJ would like to thank: Rob Hynes, Manuel Torres, Frank Verbunt for useful discussions, Lucas Ellerbroek, Andrea Modigliani and Wolfram Freudling for the help with {\it X-Shooter} data reduction, Tom Marsh for the use of {\sc molly} and LTE emission line models, Keith Horne for the use of the eclipse mapping code. PGJ acknowledges support from a VIDI grant from the Netherlands Organisation for Scientific Research. We acknowledge the {\it RXTE/ASM} and {\it MAXI} teams for the X-ray monitoring data.



\end{document}